\documentclass[pdflatex,sn-basic]{article}

\usepackage[sc]{mathpazo} 
\usepackage{fullpage}
\usepackage[authoryear,sectionbib,sort]{natbib}
\usepackage[utf8]{inputenc}
\usepackage{lineno}
\usepackage{titlesec}
\usepackage{amsmath} 
\usepackage{xcolor}
\usepackage{natbib}
\usepackage{graphicx}
\usepackage{pifont}
\usepackage{float}
\usepackage{tikz}
\def\checkmark{\tikz\fill[scale=0.8](0,.35) -- (.25,0) -- (1,.7) -- (.25,.15) -- cycle;} 
\usepackage{array}
\newcolumntype{P}[1]{>{\centering\arraybackslash}p{#1}}
\usepackage{}
\titleformat{\section}[block]{\Large\bfseries\filcenter}{\thesection}{1em}{}
\titleformat{\subsection}[block]{\Large\itshape\filcenter}{\thesubsection}{1em}{}
\titleformat{\subsubsection}[block]{\large\itshape}{\thesubsubsection}{1em}{}
\titleformat{\paragraph}[runin]{\itshape}{\theparagraph}{1em}{}[. ]\renewcommand{\refname}{Literature Cited}

\title{Rethinking tipping points in spatial ecosystems }

\author{Swarnendu Banerjee$^{1,2,3,\ast}$,
Mara Baudena$^{4,5,1}$, 
Paul Carter$^{6}$,
Robbin Bastiaansen$^{7,8}$\\
Arjen Doelman$^{9}$,
Max Rietkerk$^{1}$}

\date{}

\begin{document}

\maketitle

\noindent{} 1. Copernicus Institute of Sustainable Development, Utrecht University, 3508 TC, Utrecht, The Netherlands;

\noindent{} 2. Centre for Complex Systems Studies, Utrecht University, Utrecht, The Netherlands;

\noindent{} 3. Dutch Institute for Emergent Phenomena, Institute for Biodiversity and Ecosystem Dynamics, University of Amsterdam, 1090 GE Amsterdam, The Netherlands.

\noindent{} 4. National Research Council of Italy, Institute of Atmospheric Sciences and Climate (CNR-ISAC), 10133 Torino, Italy

\noindent{} 5. National Biodiversity Future Center, 90133 Palermo, Italy 

\noindent{} 6. Department of Mathematics, University of California Irvine, Irvine, USA

\noindent{} 7. Mathematical Institute, Utrecht University, 3508 TC, Utrecht, The Netherlands

\noindent{} 8. Department of Physics, Institute for Marine and Atmospheric Research Utrecht, Utrecht University, Utrecht, The Netherlands

\noindent{} 9. Mathematical Institute, Leiden University, 2300 RA, Leiden, Netherlands.

\noindent{} $\ast$ Corresponding author; e-mail: swarnendubanerjee92@gmail.com

\bigskip

\textit{Manuscript elements}: Figure~1, figure~2, figure~3, figure~4, figure~5, figure~6, figure~7, Box~1, supplemental PDF.

\bigskip




\bigskip



\newpage{}

\section*{Abstract}

The theory of alternative stable states and tipping points has garnered substantial attention in the last several decades. It predicts potential critical transitions from one ecosystem state to a completely different state under increasing environmental stress. However, typically, ecosystem models that predict tipping do not resolve space explicitly. Ecosystems being inherently spatial, it is important to understand the effects of spatial processes. In fact, it has been argued that spatial dynamics can actually help ecosystems evade tipping. Here, using a dryland and a savanna-forest model as example systems, we provide a synthesis of several mechanisms by which spatial processes can change our predictions of tipping in ecosystems. We show that self-organized Turing patterns can emerge in drylands that help evade tipping, but that (non-Turing) patterns driven by environmental heterogeneity are key to evasion of tipping in humid savannas. Since the ecological interactions driving the dynamics of these ecosystems differ from each other, we suggest that tipping evasion mechanisms in ecosystems may be connected to the key ecological interactions in a system. This highlights the need for further research into the link between the two in order to formulate better strategies to make ecosystems resilient to global change.

\newpage{}

\section*{Introduction}

\begin{figure}[h!]
 \begin{center}
{\includegraphics[width=1\textwidth]{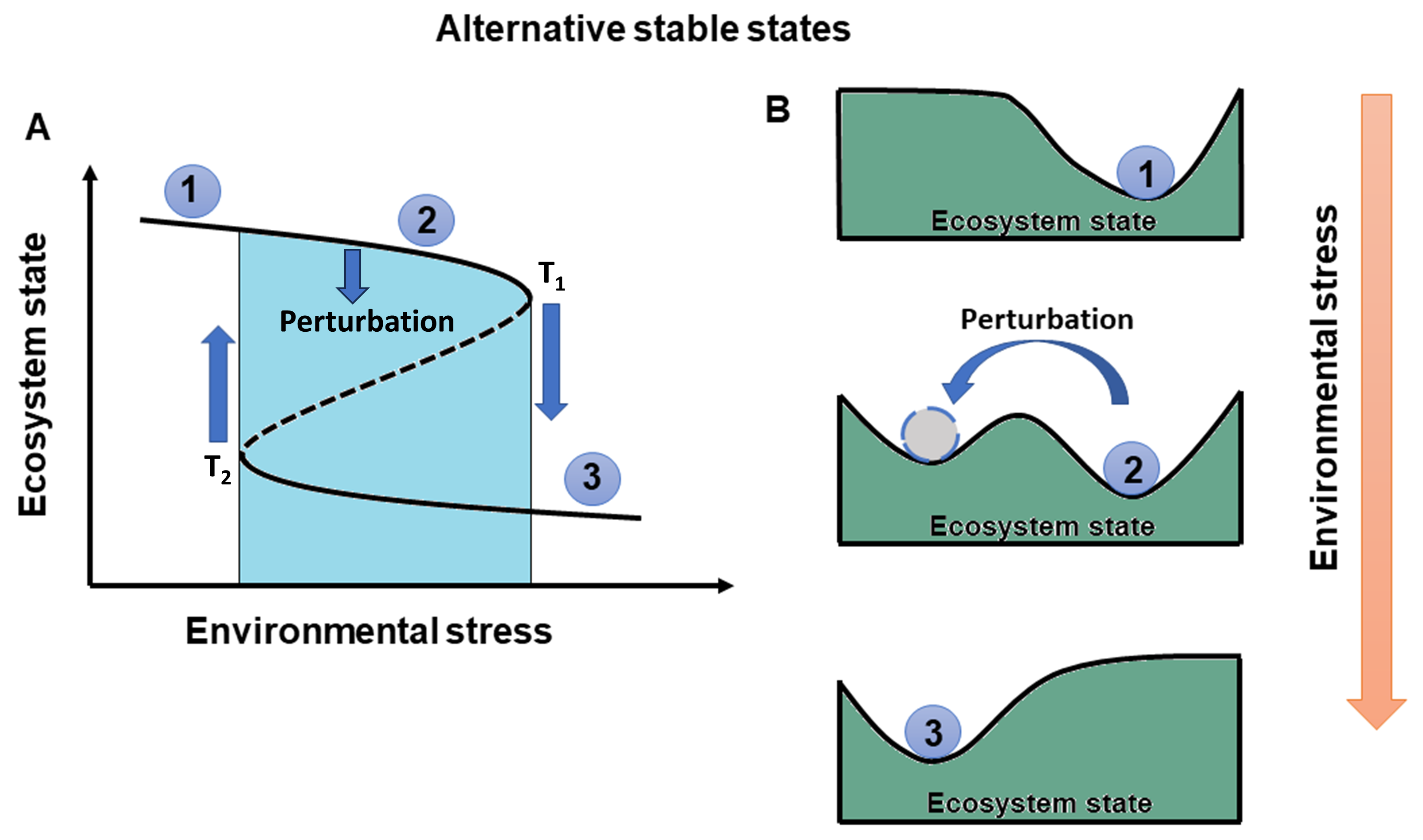}}
 \end{center}
 \caption{{\small A. Classical theory of tipping points: with increasing environmental stress, the ecosystem which is in state \textcircled{1} undergoes a critical transition to the alternative stable state \textcircled{3} at $T_1$ and the stress has to decrease below $T_2$ before the ecosystem returns to the original state via another critical transition. When the ecosystem is in state \textcircled{2}, a perturbation which pushes the system beyond the unstable state could also lead to transition to the alternative stable state. The blue shaded region indicates the environmental conditions under which alternative stable states exist. The solid and dotted lines represent stable and unstable ecosystem states respectively. B. Changes in stability landscape with increasing environmental stress (from top to bottom). Numbered balls in A and B correspond to the same ecosystem state for the same environmental stress levels.} }
 \label{Fig1}
 \end{figure}

The idea that alternative stable states may exist in ecosystems has prevailed in ecological literature for almost half a century now \citep{holling1973resilience, may1977thresholds, noy1975stability}. This directly relates to the theory of tipping points, which contributes to our understanding of ecosystem functioning \citep{scheffer2001catastrophic, scheffer2009early, van2016you}. With increasing stress, an ecosystem persists in one of its stable states until a critical point (also called a tipping point) is reached and a critical transition (or tipping) to another stable state occurs (Fig. \ref{Fig1}A). Once the system has crossed a tipping point, decreasing stress to the original value does not restore the system to the original state. This phenomenon of history dependence is called hysteresis. The ecological significance of such a phenomenon is that a tiny change in environmental conditions can result in irreversible change of the system to another state with completely different characteristics. 

The theory of alternative stable state has been used by ecologists to explain, for instance, the process of clear lakes rapidly turning turbid \citep{banerjee2021chemical, scheffer1993alternative}, desertification of vegetated lands \citep{rietkerk1997alternate, rietkerk1997site} and the co-occurrence of tropical forests and savannas for the same climatic conditions \citep{staver2012integrating, touboul2018complex}. Positive feedbacks drive ecosystems to undergo such critical transitions \citep{deangelis1986positive, van2016you}. For instance, in the case of shallow lakes, nutrient overloading causes water turbidity, which leads to decrease in density of macrophyte plants, which in turn leads to even lesser nutrient uptake and so more turbidity \citep{scheffer1993alternative}. Similarly, in the case of dryland ecosystems, a decrease in vegetation reduces water infiltration into the soil, which in turn leads to further vegetation decrease \citep{rietkerk1997alternate}. In recent years, the changing global environment due to climate change led to a renewed resurgence of these ideas whereby the concept of critical transitions has been linked to ecosystems as well as climate system elements \citep{lenton2008tipping}.

Alternative states in ecosystems are often explained by the ball in a landscape analogy (Fig. \ref{Fig1}B). Here, a ball represents the current ecosystem state and the landscape represents its stability properties (its ‘stability landscape’) which depends on the environmental conditions or parameters. When there is only one stable state, the landscape has only one valley where the ball rests. Increasing environmental stress or changing model parameters alters the shape of the landscape. If changing environment drives the system to an alternative ecosystem state zone, two valleys emerge, each representing a stable state. Then the peak in between these two valleys represents a third equilibrium that is unstable. For further increase in stress levels, the landscape may change such that the original valley disappears and the ball rolls to the alternative valley. The mathematical phenomenon by which valleys form or disappear in the landscape is known as a bifurcation. This is the underlying cause for critical transition or (bifurcation-induced) tipping. A majority of the literature on tipping or abrupt transition has been focused around fold bifurcations where two equilibria, one stable and one unstable meet and disappear. Tipping can also occur before the bifurcation point (tipping point) if a perturbation or a disturbance pushes the system beyond the unstable equilibrium. This is commonly referred to as noise induced tipping or N-tipping in the literature \citep{ashwin2012tipping}. 

While the notion of tipping or critical transitions in ecosystems has been typically understood by simple non-spatial models, spatial processes such as dispersal and lateral flows are ubiquitous in nature. Taking into consideration such mechanisms can lead to predictions of spatially patterned ecosystem states \citep{goel2020dispersal,siteur2014beyond,zelnik2017desertification}. The consequences of these patterned states on ecosystem transitions have been studied using dryland models \citep{bel2012gradual, siteur2014beyond, zelnik2017desertification, zelnik2018regime}, and it has been argued that patterns can help ecosystems evade tipping \citep{rietkerk2021evasion}. Nevertheless, the literature lacks a comprehensive synthesis of the different kinds spatial mechanisms which challenge the notion of large scale tipping as predicted by non-spatial models. The goal of this synthesis is to address this gap while recognizing that the ecological interactions which are at play in different ecosystems determine the spatial mechanism which is relevant there. For instance, in savannas, the key ecological processes which drive the system differ along a rainfall gradient. At the humid end, fire plays an important role while at the dry end, water limitation drives the system dynamics \citep{holdo2023linking}. Other interactions such as grazing might be more relevant to the savanna dynamics in regions with intermediate rainfall. In this paper we propose that these governing interactions might play a role in determining the tipping evasion mechanisms that is relevant for a particular ecosystem.

We have chosen drylands and the humid savanna-forest transition zone as archetypical ecosystems for this synthesis because of three main reasons. First, model predictions and observations suggest that humid savannas and tropical forests may exhibit tipping at the boundary. For instance, the savanna-forest transition zone could be bimodal, i.e., for the same climatic conditions, both savanna and tropical forests are observed \citep{aleman2020floristic, dantas2016disturbance, d2018not, hirota2011,staver2011global}. Further, in drylands, i.e. ecosystems with arid to semi-arid conditions, it is predicted that desertification could occur via critical transitions or tipping \citep{rietkerk1997alternate, rietkerk1997site}. Second, at both savanna-forest transition zone and drylands, spatial patterns have been observed \citep{groen2007spatial}. At the humid savanna-forest boundary, the role of spatial scale and heterogeneity has been stressed upon recently \citep{aleman2018spatial, staver2018prediction, wuyts2017amazonian}. In semi-arid ecosystems, various kinds of vegetation patterns such as gaps, labyrinth, stripes and spots are well documented in the literature \citep{deblauwe2008global, rietkerk2002self, rietkerk1997alternate}, which highlights the importance of spatial processes. Third, as mentioned earlier, the ecological interactions which are key to driving the dynamics in these ecosystems are different from each other. In drylands, vegetation biomass increases soil water infiltration thus leading to more vegetation, which is a positive feedback. In humid savanna-forest boundary, the positive feedback is due to interaction between the two vegetation types. The savanna grasses  act as fuel to fire and harm the forest trees, while  the forest trees create shade over the savanna vegetation, which is shade intolerant. Together, these negative effects lead to an overall positive feedback reinforcing savanna and forest states above certain thresholds.

In this paper, we represent drylands and humid savanna-forest boundary with two conceptual models. We first show that the non-spatial versions indeed exhibit tipping behavior which can be attributed to the positive feedback in the system. Subsequently, we incorporate spatial dynamics in these models and demonstrate the various tipping evasion mechanisms due to diverse dynamics, such as interacting spatial feedbacks, spatial heterogeneity and behavior of spatial boundaries (also called fronts or interfaces) between alternative stable states. We identify the mechanisms that might be more relevant for each biome (see Table \ref{table-1}) and provide an intuitive understanding of their connection with the ecological interactions at play. Finally, we discuss the implications of these developments and delineate future perspectives for studies contributing towards consequences of global change.

\section*{Modelling drylands and humid savannas}

Drylands and humid savannas have both been modelled in the past with different levels of complexity, and the non-spatial models have typically reported tipping behaviour \citep{rietkerk1997alternate, staver2012integrating}. The goal of this synthesis is to layout the various mechanisms of tipping evasion which the systems might demonstrate when spatial processes are considered. Here, we restrict our scope to discussing spatial dynamics with the help of reaction diffusion framework consisting of a reaction part which describes the local dynamics and a diffusion part which describes the spatial transport (but see \cite{kefi2007spatial, wuyts2023emergent} for examples of spatial cellular automata models).  Such models are important in the context of spatial dynamics because they are mathematically tractable and commonly used in ecology . Further, a relatively simple class of such models can exhibit rich patterns \citep{klausmeier1999regular, rietkerk2002self, wuyts2017amazonian,zelnik2015gradual}, which are comparable to what is observed in real ecosystems. For the purpose of this paper, we use models with two interacting state variables. We choose this level of complexity for several reasons. While single component models, which have been used earlier to model these ecosystems \citep{bastiaansen2022fragmented, goel2020dispersal}, are easier to analyze, they do not accommodate the richness and variety of behaviors that may be possible due to spatial effects. Although, notably, incorporating non-local coupling may lead to a variety of spatial patterns in even single component models of drylands \citep{borgogno2009mathematical, escaff2015localized,fernandez2014strong,lefever1997origin} and savannas \citep{martinez2013spatial}, we choose to use two components, which allows us to model explicitly the key ecological interactions relevant to each of the ecosystems. We refrain from incorporating more than two state variables for the ease of analysis, while this level of complexity is also prevalent in the literature. \citep{staver2012integrating, gilad2004ecosystem, rietkerk2002self, wuyts2017amazonian}. The savanna-forest boundary has been modelled earlier using vegetation cover as a state variable, considering space implicitly \citep{staver2012integrating, touboul2018complex, wuyts2017amazonian, wuyts2019tropical}. Since it is common to model vegetation biomass for dryland ecosystems, for better comparison, we choose to study a vegetation biomass model of the savanna inspired from earlier works by \cite{beckage2009vegetation} and \cite{yatat2018spatially} which is extended to include explicit spatial dynamics. Using models of similar complexity for both cases, allows us to compare the two models and generate new hypotheses about the differences in their behaviour upon spatial extension. 

\subsection*{Non-spatial dynamics}

\subsubsection*{Dryland ecosystem}

To represent the vegetation dynamics in drylands, we consider a modified version of the commonly used Klausmeier model \citep{bastiaansen2019stable, eigentler2021species, klausmeier1999regular}. Although this model has been largely used in the spatial context, we start with a non-spatial variant here to highlight the differences in behavior predicted from the spatial and the non-spatial version. The non-spatial (i.e., “reaction”) part of the model (hereafter referred to as “dryland model”), which describes the water-plant interactions in drylands, is given as follows:

\begin{eqnarray}
\begin{array}{llll}
\displaystyle \frac{dW}{d\tau}= q-lW-rWV^2  \\\\

\displaystyle \frac{dV}{d\tau}=rjWV^2 (1-V/K)-dV \\\\

\end{array}
\end{eqnarray}

Here, $W$ represents soil water and $V$ represents vegetation biomass, $q$ represents rainfall and $lW$ describes the water evaporation rate. The uptake rate of water by plants is given by $(rWV)V$, where $rWV$ can be understood as the product of two linear functions, $rW$ and $V$. The former is the functional response of plants to water while the later describes how the water infiltration increases with plants \citep{klausmeier1999regular}. An increase in vegetation biomass leads to increased water uptake, which in turn leads to further increase in vegetation, thus forming a positive feedback. So, the vegetation functions as an activator to itself. The conversion of water into biomass is assumed to only occur until the vegetation reaches a carrying capacity $K$. Hence, it can be expressed as $j(1-V/K)$ which is a decreasing function of $V$ until it reaches zero at the carrying capacity, $V=K$. $d$ represents natural mortality of the vegetation.
The model can then be non-dimensionalized (see Supplementary material Section B.1) as follows:

\begin{eqnarray}
\begin{array}{llll}
\displaystyle \frac{dw}{dt}= p-w-wv^2 = F(w,v) \\\\

\displaystyle \frac{dv}{dt}=wv^2 \left(1-\frac{v}{k}\right)-uv=G(w,v) \\\\

\end{array}
\end{eqnarray}

where $w$ and $v$ are the dimensionless state variables and $p$, $u$, and $k$ are positive parameters. Note that the interaction between the two components in this model, vegetation and water, can be understood with the help of the partial derivatives, $F_v(w,v)$ and $G_w(w,v)$. The signs of these quantities show how one of the components influence the rate of the change of the other. Here, $F_v\leq0$ due to the uptake of water by vegetation and $G_w\geq0$ as vegetation grows with the help of water (for vegetation below carrying capacity). So the water acts as a substrate for vegetation and the latter also activates itself. Such a system can be referred to as an activator-substrate system \citep {koch1994biological, meinhardt2012turing}. Since water limitation drives the system dynamics, the rainfall parameter, $q$, in the first model, is an appropriate choice for the bifurcation parameter. In the dimensionless model, this corresponds to $p$. All parameters are adapted from earlier literature (see Supporting information, section 2.5). 

A bifurcation diagram of the model shows that it possesses two alternative stable states that exist for a certain parameter range, namely (i) a bare soil state and (ii) a vegetated state (Fig. \ref{Fig2}A).  As rainfall decreases, vegetation biomass decreases slowly until it reaches a threshold where vegetation can no longer persist, and the system undergoes a critical shift to the bare soil state. This shift is completely irreversible as on increasing rainfall, the system can no longer go back to the vegetated state. Note that in other models of critical transitions in dryland, it may be possible to reach the vegetated state from a bare soil state on increasing rainfall \citep{rietkerk2002self} but this is inconsequential to the purpose of this study. 

\subsubsection*{Savanna-forest boundary}

To represent vegetation dynamics at the humid savanna-forest boundary, we consider two plant functional types: grasses, which fundamentally define savannas vegetation \citep{lehmann2011deciphering, parr2014tropical}, and forest trees. For simplicity, we do not explicitly consider savanna trees, but we cluster them together with grasses into the savanna vegetation variable, since the savanna trees have similar responses to shade and fire as grasses \citep{charles2018steal}. It has been established that fire plays an important role at the savanna-forest boundary \citep{dantas2016disturbance, de2013fire} and hence we take into account the effect of fire. Grasses are shade intolerant, fire resistant and have good resprouting abilities, which thus results in minimal loss and quick regrowth after fire \citep{hoffmann2012ecological}. Conversely, forest trees can tolerate shade well but can suffer large destructions in the presence of fire \citep{charles2018steal}. So, grasses act as fuel enabling fire to spread, thus playing a significant role in fire mediated destruction of the forest trees \citep{lehmann2011deciphering}. Conversely, tropical forest trees closes canopy and has a negative effect on the shade intolerant grasses. The combination of these two negative feedback leads to an overall positive feedback which is at play at the savanna-forest boundary. 

We denote savanna biomass density by $S (kg m^{-2})$ and forest biomass by $F (kg m^{-2})$, and model their dynamics as:

\begin{eqnarray}
\begin{array}{llll}
\displaystyle \frac{dS}{d\tau}= r_SS \left(1-\frac{S}{K_S} \right) -cFS-\eta f_S S^2-d_S S   \\\\

\displaystyle \frac{dF}{d\tau}=r_F F\left(1-\frac{F}{K_F} \right)- \eta f_F SF-d_F F  \\\\

\end{array}
\end{eqnarray}

In this model (hereafter referred to as "savanna-forest model"), logistic growth is considered, which allows the forest and savanna biomass to increase until their carrying capacity is reached. $r_S$ and $r_F$ are the growth rate of savanna vegetation and forest trees respectively. $K_S$ and $K_F$ are their carrying capacities. $d_F$ and $d_S$ are the removal rate of the forest trees and of the savanna vegetation, resulting from factors such as natural mortality and herbivory. For convenience of parameterizing the model (see Supplementary), mortality rates have been considered separately from  the growth rates of the vegetation instead of combining them together. Savanna vegetation acts as a fuel to the fires, which occur at a frequency $\eta$. $f_S$ and $f_F$ represent the sensitivity of savanna vegetation and forest trees to fire, with $f_F > f_S$ since savanna vegetation is more fire-resistant than forest trees.  Thus, the rate at which savanna biomass is lost is proportional to the savanna plant biomass and can be modeled as -$\eta f_S S$. Further, fire also causes destruction and loss of forest biomass and the rate at which such loss occurs is modeled with -$\eta f_F S$. Although it is common to use a sigmoidal function for the effect of fire on forest cover \citep{staver2012integrating, goel2020dispersal}, we consider here, a linear function for simplicity \citep{beckage2009vegetation}. Since tropical forests are dense and wet, they are unlikely to sustain fire by themselves \citep{hoffmann2009tree, cochrane2003fire} and hence it is reasonable to assume that fire only impact forest trees in the presence of savanna biomass which acts as a fuel. Forest trees with closed canopy have a negative influence on the shade-intolerant savanna vegetation which is modeled with $-cFS$. $\tau$ is time expressed in years. The dimensionless version of the model can be expressed as follows (see Supplementary material Section A.1 for details): 

\begin{eqnarray}
\begin{array}{llll}
\displaystyle \frac{ds}{dt}= s(1-s)-bfs-ns =H(s,f)  \\\\

\displaystyle \frac{df}{dt}=\mu f (1-f)- asf-mf=I(s,f)  \\\\

\end{array}
\end{eqnarray}

\begin{figure}[H]
 \begin{center}
{\includegraphics[width=1\textwidth]{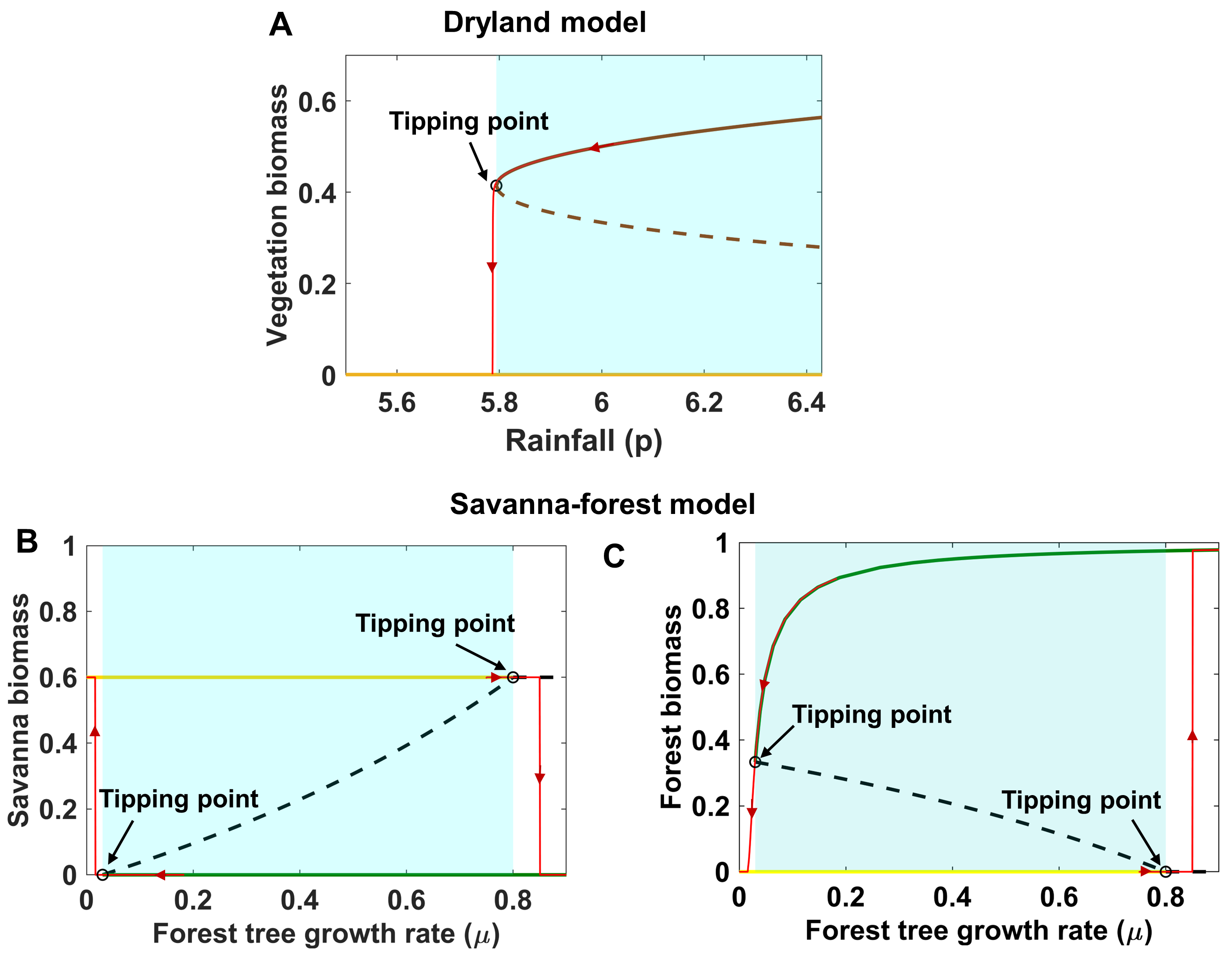}}
 \end{center}
 \caption{{\small Bifurcation diagrams of the two dimensionless non-spatial models. (A) Dryland model (Equation 2), vegetation biomass as a function of rainfall (p). The dark brown line represents the vegetated state and the light brown line represent the barren state. Parameter values: $u=1.2, k=1$. (B,C) Savanna-forest model (Equation 4), savanna (B) and forest (C) biomass as a function of forest tree growth rate ($\mu$). The green lines represent forest-only state and the yellow lines represent savanna-only state. The solid lines represent stable equilibria while the dashed lines represent unstable equilibria. For clarity, other biologically non-feasible equilibria are not shown here (see Supplementary material Section A.5, Fig S1,  for the complete bifurcation diagram). Parameter values: $a=1.3, b=1.8, m=0.02, n=0.4$. The red lines denote time series simulations with the respective equilibria as initial conditions and a gradual decrease in bifurcation parameters, i.e., $dp/dt=-0.0005$ (dryland) and $d\mu/dt=0.0005$ (savanna-forest).}}
 \label{Fig2}
 \end{figure}

Here, $s$ and $f$ are the dimensionless state variables and $\mu$, $a$, $b$, $m$, $n$ are the dimensionless parameters of the model. Here, the signs of the partial derivatives, $H_f(s,f)$ and $I_s(s,f)$ are both negative, i.e., $H_f(s,f)\leq0$ and $I_s(s,f)\leq0$. This is essentially due to the negative effect of the forest trees on the savanna vegetation and vice-versa. In order to explore the equilibrium properties of the savanna-forest model, we carry out a bifurcation analysis of the non-spatial model (Eqs. (2)). Water availability, mostly due to rainfall, is expected to be an important bifurcation parameter as tropical forests ultimately dominate when water availability is very large, while savannas are observed at lower mean annual rainfall (e.g \citep{lehmann2011deciphering}) . As we are not modelling water availability explicitly, we represent, the changing climatic condition along the rainfall gradient in this transition zone by changing the forest tree growth rate ($r_F$ in the dimensional model (1)). This is justified as precipitation increases the forest tree growth rate, while it has negligible impact on the growth of the savanna vegetation in mesic and humid savannas, where  the prevailing humid conditions are sufficient so that the grasses are not limited by water \citep{d2018not, holdo2023linking}. Since $\mu=r_F/r_S$   (see Supplementary material Section A.1) is the nondimensional forest tree growth rate, we choose $\mu$ as the bifurcation parameter (Fig. \ref{Fig2} B,C) and for brevity we name it “forest tree growth rate”. The choices of the parameters are motivated from ecological literature (see Supplementary material Section A.5). 

In an intermediate range of forest tree growth rate, $\mu$, the forest and savanna states exist as alternative stable states. On decreasing $\mu$, if the system is in the forest state it undergoes an abrupt transition to the savanna state. On increasing $\mu$, if the system is in the savanna state, it tips to the forest state. The tipping behavior demonstrated in this model is due to two transcritical bifurcations. Although transcritical bifurcation are often associated with smooth transitions from one stable branch to another \citep{kefi2013early}, the clearly discontinuous nature of the ecologically feasible equilibria in Fig. \ref{Fig2}, results in the system to tip (see Supplementary material Section A.2, A.3 for analytical expression of the steady states and their stability conditions and Section A.5, Fig S1 for the complete bifurcation diagram). This is noteworthy because in many other similar models including the dryland model which we described previously, tipping is encountered through fold bifurcations (Fig. 1). \citep{staver2012integrating, van2015resilience}. 

\subsection*{Spatially extended ecosystems}

We now spatially extend these two models and demonstrate the different mechanisms by which the predictions of alternative stable state theory change when incorporating spatial effects. We consider the evolution of the two dynamical variables in each model, $\begin{pmatrix}w \\ v\end{pmatrix}$ in case of the dryland model, and  $\begin{pmatrix}s \\ f\end{pmatrix}$ in case of the savanna-forest model, on an unbounded spatial domain with coordinates $(x,y)\in \mathbb{R}^2$  (or only $x\in \mathbb{R}$ in case of one dimension). This means that in addition to the local dynamics, spatial transport between locations now also play a role. The simplest way to implement spatial transport is to add (linear) diffusion (but see \cite{van2013rise} for example of non-linear diffusion). In the case of plants, this represents dispersal. So, the new spatial models, also known as reaction-diffusion equations, describing the vegetation dynamics in savanna-forest boundary and drylands are given by the partial differential equations which are as follows:\\

Dryland model:
\begin{eqnarray} 
\begin{array}{llll}
\displaystyle \frac{\partial w}{\partial t} = p- w - wv^2 + \delta_w \left(\frac{\partial^2w}{\partial x^2}+\frac{\partial^2w}{\partial y^2}\right)\\\\

\displaystyle \frac{\partial v}{\partial t}= \displaystyle wv^2\left(1-\frac{v}{k}\right)- uv + \left(\frac{\partial^2v}{\partial x^2}+\frac{\partial^2v}{\partial y^2}\right)\\
\label{dryland} 
\end{array}
\end{eqnarray}

Savanna-forest model:
\begin{eqnarray} 
\begin{array}{llll}
\displaystyle \frac{\partial s}{\partial t}= \displaystyle s (1-s)-bfs-ns + \left(\frac{\partial^2s}{\partial x^2} + \frac{\partial^2s}{\partial y^2}\right) \\\\

\displaystyle \frac{\partial f}{\partial t}= \mu f\left(1-f\right)- asf - m f + \delta \left(\frac{\partial^2f}{\partial x^2}+\frac{\partial^2f}{\partial y^2}\right)\\

\label{forest_dimensionless} 
\end{array}
\end{eqnarray}

In Eqs. 5, $\delta_w$ represents the ratio of the diffusion coefficient of water and dryland vegetation and in Eqs. 6, $\delta$ represents the ratio of the diffusion coefficient of the forest and savanna vegetation (see Supplementary material Section A.1. and B.1 for non-dimensionalization of the full spatial models).

\section*{Mechanisms of tipping evasion}

In this section, we discuss with the help of above examples, the various mechanisms of tipping evasion that may be observed when models with alternative stable states are spatially extended. We discuss regular `Turing' pattern formation which is common in many models of drylands. Furthermore, patterns could also be formed due to local disturbances – which we discuss in the context of both savanna-forest boundary and drylands and then compare the two cases. We also describe the role of spatially heterogeneous environment and front instabilities in evading tipping from one alternative ecosystem state to another. All simulations in have been carried out using MATLAB.

\subsection*{Turing before tipping}

\begin{figure}[H]
 \begin{center}
{\includegraphics[width=0.8\textwidth]{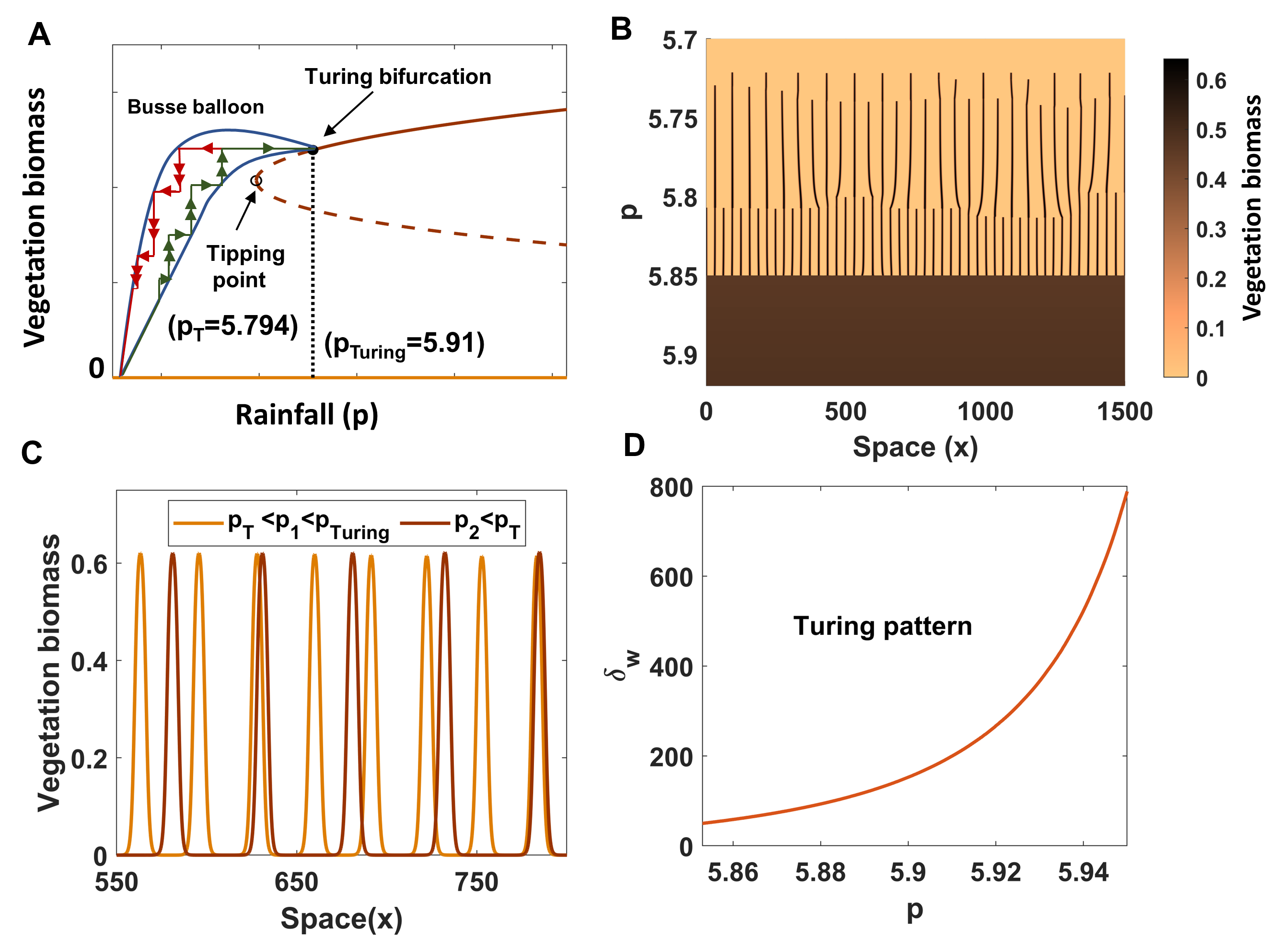}}
 \end{center}
 \caption{{\small Turing before tipping in the dryland model. (A) Turing bifurcation occurs for larger values of rainfall ($p_{Turing}$) compared to the tipping point ($p_T$) in the dryland model. The ecosystem could then persist in patterned states beyond the tipping point within the Busse balloon (a conceptual diagram is presented). With decreasing rainfall, the patterned state retains its wavelength until the edge of the Busse balloon (minor ecosystem adjustments are indicated by single arrows) where the ecosystem could shift (indicated by double arrows) to a pattern of different wavelength. There could be multiple such ecosystem shifts instead of a single catastrophic state shift when the rainfall parameter, $p$, changes gradually over time. Pathway for ecosystem degradation is indicated by the red lines and for ecosystem recovery is indicated by the green lines. The dark brown line (solid: stable; dashed: unstable) represents the vegetated state and the light brown line represent the barren state. (B) Vegetation biomass across one-dimensional spatial domain on decreasing the rainfall parameter gradually, i.e., $dp/dt=-1.25 \times 10^{-6}$, with randomly perturbed homogeneous vegetation state for $p>p_{Turing}$ as initial condition. (C) Patterns of different wavelengths can be observed for $p_1=5.825>p_T$ and for $p_2=5.7625<p_T$ . (D) The region in $p- \delta_w$ parameter space where Turing patterns can be observed. Parameter values: $\delta_w=200$, other parameter values are the same as in Fig. \ref{Fig2}.}}
 \label{Fig3}
 \end{figure}

Spatially uniform equilibria in models with two or more components may become unstable to spatially heterogeneous perturbations. This is the most well-known and largely studied Turing bifurcation and results in regular spatial patterns \citep{hillerislambers2001vegetation, turing1952chemical}. On investigating the models (5) and (6) for Turing bifurcations, we see that the savanna-forest model does not exhibit such instability (for details see Supplementary material Section A.4). So, if a spatial domain is homogeneously covered with either the forest or the savanna state, then changing $\mu$ may lead to tipping of the whole spatial domain to the alternative stable state. However, for the dryland model, the onset of Turing patterns could be seen easily (Fig. \ref{Fig3} and Supplementary material Section B.4, Fig. S2) and patterns observed in many real drylands are often explained by this phenomenon \citep{rietkerk2002self, rietkerk2008regular}. Ecologically, such pattern formation can be explained by a combination of short-range facilitation and long-range inhibition due to vegetation. Vegetation patches increase water infiltration, which in turn increases vegetation locally thus resulting in a positive feedback. Water taken up by the vegetation, leads to diffusion of water from nearby areas to the vegetated patch. Hence, there is less water further away from the patch, leading to decrease in vegetation which is a negative feedback. This feedback, which changes from positive to negative with increasing distance, is known as scale dependent feedback, and it explains regular spatial pattern formation \citep{rietkerk2008regular}. Earlier studies have shown this pattern formation is typical for systems which exhibit activator-substrate mechanism \citep{meinhardt2012turing, koch1994biological}, drylands being one such example. At the savanna-forest boundary, the positive feedback is due to a reciprocal negative effect of each vegetation type. In the model, this is represented by the first order partial derivatives, $H_f$ and $I_s$ which are both negative. As such, there is no local facilitation or scale dependent feedback which could lead to formation of patches of one type of vegetation. This explains the absence of Turing patterns in the savannas. It is important to note that irrespective of the non-linearities involved in how savanna interacts with the forest and vice versa, no Turing pattern is possible in such a model where both the state variables act as inhibitor for each other (See Supplementary Section A.6). 

Notably, the Turing bifurcation exhibited by the dryland model occurs before the threshold of environmental pressure where tipping is predicted to occur in the non-spatial version of the model (Fig. \ref{Fig3}, Supplementary material Section B.4 and Fig S2). This phenomenon has been referred to as “Turing before tipping” \citep{rietkerk2021evasion}. So, essentially this implies that, when considering spatial dynamics, on decreasing rainfall there is onset of pattern formation before the system reaches the tipping point. Recent developments suggest that the behavior and stability of such patterns determines whether the ecosystem will undergo a critical transition to an alternative stable state. On changing environmental conditions, these patterns remain stable for a large range after which they typically adapt to a new wavelength. This mechanism leads to small transitions between patterns of different wavelength, which help ecosystems evade large critical transitions or tipping as predicted by non-spatial models (See Fig. \ref{Fig3}B ). Also, it has been shown for similar dryland models that for a particular environmental condition, there could be multiple stable patterns of different wavelengths. So, it is possible to define a region in the parameter wave-number (number of patches per unit distance) space, also called “Busse balloon” comprising of all the stable patterns \citep{bastiaansen2018multistability, siteur2014beyond} and the pattern adaptation occurs as the parameter reaches the boundary of this region. We do not specifically define the Busse balloon for this particular model, we present a conceptual diagram in Fig. 3A, of the general mechanism by which such a stability region helps evade tipping. While it is easy to prescribe the conditions when an activator-substrate system can lead to Turing patterns, for instance, Fig 3D provides these conditions for our model in the $p-\delta_w$ parametric space, it is not straightforward to say the system will always evade tipping via this mechanism. This will depend on the stability region of the pattern or the shape of the Busse balloon and hence a case by case investigation is recommended. 


\begin{figure}[H]
	\centering
	\fbox{
		\begin{minipage}{1.0\textwidth} 
			\textbf{Box 1: Definitions} \\

                 \textbf{Activator-substrate system:} A two component system in which the activator is a self-enhancing component, while the substrate is depeleted by the activator. In the dryland model, the vegetation is the activator while water is the substrate. \\
                \textbf{Busse balloon:} A mathematical structure in the parameter-wave number space which comprise  all the stable patterned states.\\
                \textbf{Coexistence states:} Two alternative stable states positioned next to each other in a landscape.\\
                \textbf{Environmental heterogeneity:} The spatial variation of environment within a landscape (generally given by a parameter). \\
                \textbf{Front:} The spatial interface connecting two ecosystem states (typically in a parameter regime where both are stable) \\
                \textbf{Localized disturbances:} Perturbations which push a part of the landscape to the alternative stable state. \\
                \textbf{Maxwell point:} The parameter value at which front velocity is zero, i.e., neither states invades the other.\\
                \textbf{Potential systems:} A system that describes the time evolution of a state variable $v$, is a potential systems if it is possible to define a function $U(v)$  that always decreases with time. \\
                \textbf{Scale dependent feedback:} A mechanism where there is short range facilitation and long range inhibition, e.g., water-vegetation feedback in the dryland. \\
                \textbf{Tipping point:} The point at which there is abrupt transition to the alternative stable state in the non-spatial model. \\
                \textbf{Tipping evasion:} Avoidance of abrupt transition of the ecosystem from one stable state to its alternative stable state. \\
                \textbf{Turing before tipping:} A mechanism of tipping evasion where there is onset of pattern formation as a result of a Turing bifurcation before the tipping point is reached along an axis of increasing stress.

		\end{minipage}
	}
\end{figure}

\subsection*{Spatially localized disturbances}

Although scale-dependent feedbacks (or Turing bifurcations) are probably the most well studied mechanism of pattern formation in ecosystems, they are not the only mechanism by which ecosystems can evade tipping. In models exhibiting alternative stable states, a range of environmental conditions (parameters) exists where the whole domain can be in one of the two possible states. In such an environment, spatially localized disturbances to the system state, such as grazing and local deforestation can push a part of the landscape to the alternative stable state. Instead of a system-wide critical transition or N-tipping (organised via a crossing of the basin boundary) predicted by the non-spatial framework, this may result in the two alternative states coexisting in space, while they are connected via a spatial boundary. Such a boundary, also called ‘front’ in the mathematical literature, propagates in one of the directions, resulting in either a gradual recovery from initial disturbance or a gradual but complete transition to the alternative stable state \citep{bel2012gradual, van2015resilience}. While such a front always exists for potential systems (see Box 1 for definition of potential systems), many two- or multi-component models representative of ecosystems are often not potential systems (see Box 1). For instance, the dryland and the savanna-forest model described above are not potential systems and so we check if such fronts exists in those cases. We can demonstrate numerically, travelling fronts for both of our models and the direction of front propagation can be explained by environmental conditions represented by model parameters. For instance, in the spatial dryland model, if we are in the bistability range and the landscape is initially divided into vegetated land and bare soil, we see that either the vegetation invades the bare soil or vice versa depending on the parameter values. The parameter $p$, representing precipitation, govern the direction of propagation of the front (see Fig. 4). For parameters corresponding to Fig. \ref{Fig2}, when $p=6.3$, the bare soil invades the vegetation and eventually covers the whole landscape and when $p=6.4$, the opposite happens. Ecologically this implies that changing climatic conditions could govern whether vegetation in a landscape will recover or there will a complete transition to bare soil in response to disturbance.

\begin{figure}[H]
 \begin{center}
{\includegraphics[width=1\textwidth]{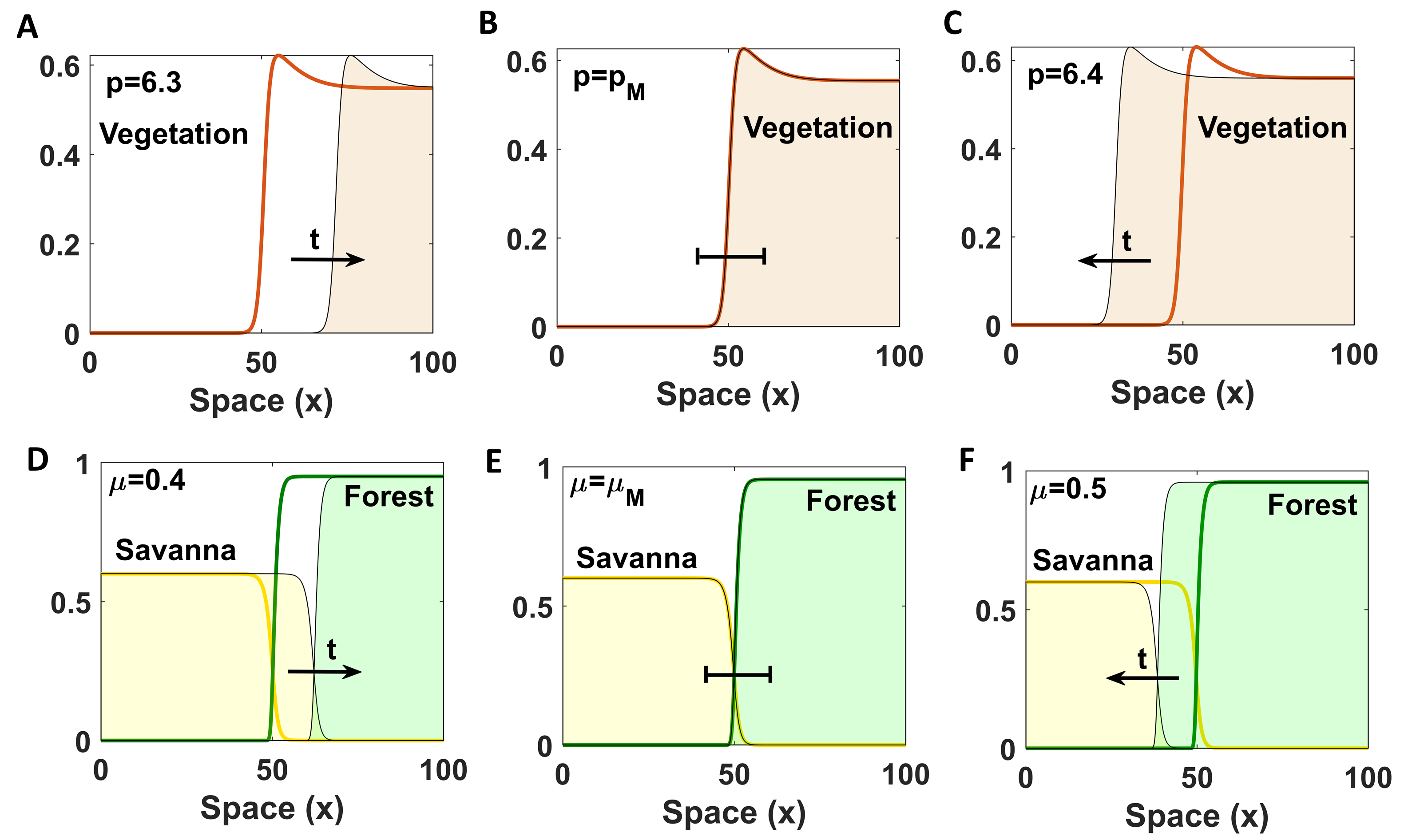}}
 \end{center}
 \caption{{\small Front propagation over time in one dimensional space. Direction of front propagation is indicated by the arrows. (A-C) Front connecting vegetation and the barren states. Vegetation invades barren land when p=6.4, and barren land invades vegetation when p=6.3. For $p=p_M (\sim 6.3513)$, the front remains approximately stationary. The brown line indicates vegetation biomass close to the initial condition where half of the spatial domain is in the barren state and the other half is in the vegetated state. Parameter values: $\delta_w=100$, the other parameters are the same as in Fig 2. (D-F) Fronts connecting the savanna and the forest states. The savanna invades the forest when $\mu=0.4<\mu_M$, and the forest invades the savanna when $\mu=0.5>\mu_M$. For $\mu=\mu_M(\sim 0.4489)$, the front remains approximately stationary. Green and yellow lines denote forest and savanna biomass respectively, close to the initial condition where half of the spatial domain is in the savanna state and the other half is in the forest state.  The shaded regions indicate their respective biomass after a time t. Parameter values: $\delta=0.01$, the other parameters are the same as in Fig. \ref{Fig2}. } }
 \label{Fig4}
 \end{figure}

Interestingly, there is a parameter value at which the front remains stationary, i.e., neither of the states invades the other, which is called “Maxwell point” \citep{pismen2006patterns} (see Box 1). At this point, in other words, the velocity of the front propagation becomes zero and thus the part of the landscape covered by each of the ecosystem states remains the same in size. The Maxwell point in the above case will be $6.3<p_M<6.4$ and can numerically be approximated to be $p_M\sim 6.3513$ (Fig \ref{Fig4}.B).  It is noteworthy that any slight deviation from that environmental condition at the Maxwell point would lead to a transition of the whole spatial domain to either of the ecosystem states, thereby exhibiting a characteristic of an unstable state.  Similar results also hold true for the savanna-forest model where smaller forest tree growth rate $(\mu=0.4)$ leads to savanna invading the forest and a larger value of the same parameter $(\mu=0.5)$ results in the movement of the front in the opposite direction (Fig. \ref{Fig4}D-F). Since forest tree growth rates are expected change along the rainfall gradient, ecologically this implies that changing climatic conditions could govern whether savanna invades the forest or vice versa. The “Maxwell point” in this case is approximately $\mu_M\sim 0.4489$ (Fig. \ref{Fig4}E).

The above-described front propagation can also lead to stable spatial patterns when multiple localized disturbances form localized domains of the alternative stable states in the landscape. This could happen when a homogeneous landscape is disrupted by, for instance, changes in land use. For example, multiple large patches of bare soil could form within an otherwise vegetated landscape in drylands. Mathematically, this results in the creation of multiple fronts whose dynamics is more intricate than that of a single front. The fronts (i.e. the edges of the patches) may move towards each other, leading the vegetation patches or the barren areas to merge, depending on the direction of front propagation \citep{bel2012gradual}. However, the presence of one front may also influence the movement of the other. In the case of two- or multicomponent systems, this can result in stationary fronts. Such multiple stationary fronts  may lead to domains of alternative stable states positioned next to each other, so they can coexist in a stable way. So, for environmental conditions where the barren state can invade the vegetated state (Fig. \ref{Fig5}A), multiple stationary fronts created via local disturbances can lead to coexistence states (Fig \ref{Fig5}B), consisting of alternating patches of vegetation and bare soil. Similar behaviours have been found in other dryland models \citep{jaibi2020existence, zelnik2018regime}. Now it is important to ask whether these coexistence states can exist beyond the tipping point predicted by the non-spatial model. In Fig. \ref{Fig5}C, for a certain choice of parameters (environmental conditions), we demonstrate that this is indeed possible for the dryland model: a coexistence state consisting of multiple fronts persists beyond the tipping point even when one of the alternative stable states giving rise to the fronts — in this case the vegetated state — ceases to exist there.

We are unaware of any general conditions that would allow for such behaviour – as such it is not straightforward to determine a relation between the functioning of an ecosystem and whether such patterns might occur. For comparison, we test whether the savanna-forest model, where the interaction between the two components is different ($H_f<0$, $I_s<0$) compared to the dryland model ($G_w>0$, $F_v<0$), shows similar behaviour. Under the same parametric condition as in Fig. \ref{Fig4}, we ran simulations starting with a patch of savanna in a landscape of forest. The savanna patch was varied between very small and almost as large as the total landscape. In all cases, the forest always invades the savanna. Further, we checked the dependency of this behaviour on the parameters $\mu$ and $\delta$. $\mu$ was randomly chosen between 0.035 and 0.75 (the bistability range being 0.03 to 0.8) and $\delta$ between 0.0025 and 0.0225
and simulations was carried out with an initial savanna patch in a forest landscape. This was repeated for 3000 times and none of the simulations resulted in a stable pattern.  In order to find stable patterns, an additional condition of environmental spatial heterogeneity is necessary, as we  discuss in the next section (see Fig 5D-E). As a preliminary test to determine whether this depends on the functional form, we also ran the simulation for a modified savanna-forest model where the interaction between forest and savanna follows a saturating function (see Supplementary Section E). This yielded the same result thus indicating that the difference in behaviour between the two ecosystems might be an outcome of the different nature of the ecological interactions in these systems. 

\subsection*{Spatial heterogeneity}

\begin{figure}[H]
 \begin{center}
{\includegraphics[width=1\textwidth]{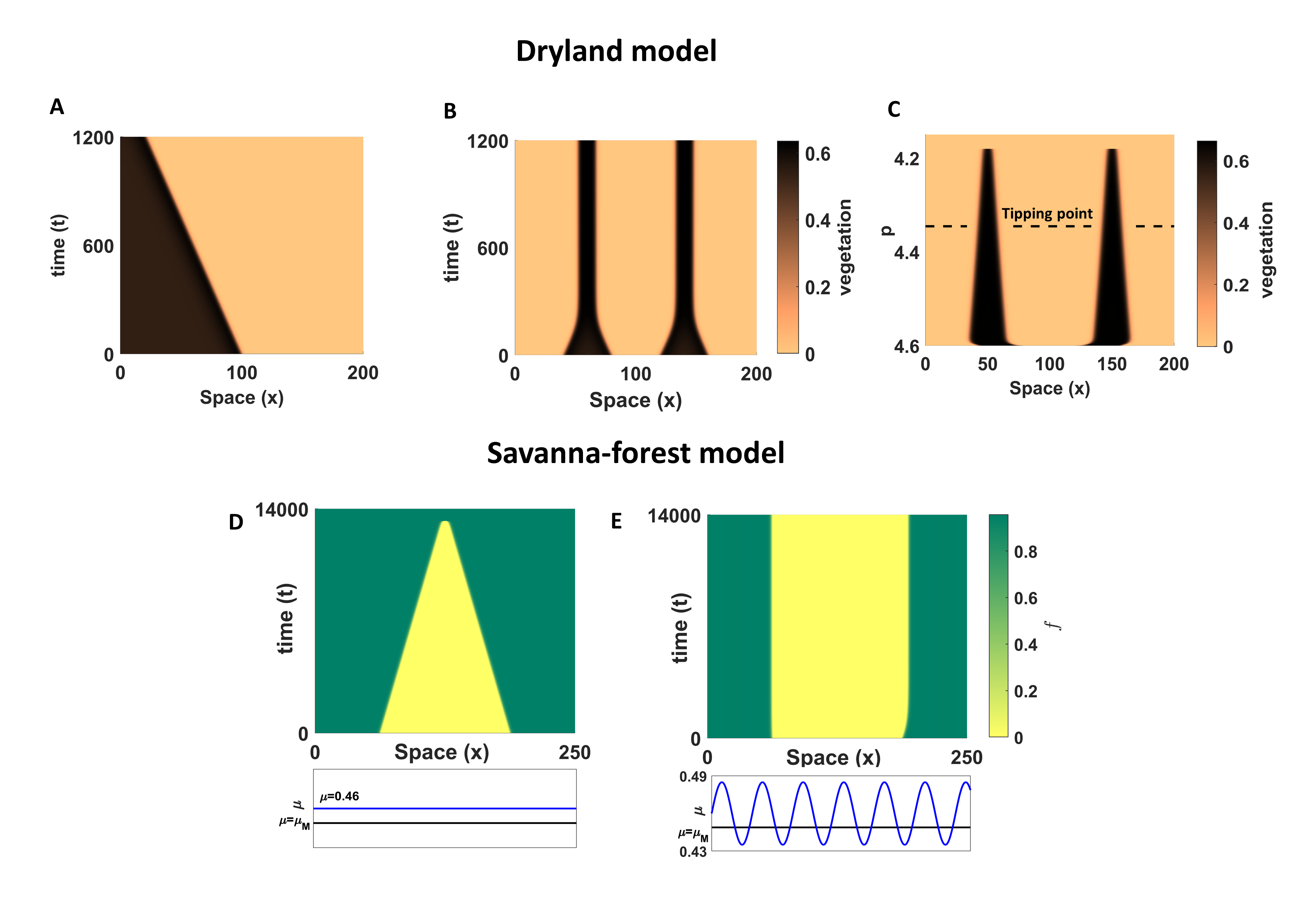}}
 \end{center}
 \caption{{\small Coexistence states. (A) Barren state invading the vegetated state in the dryland ecosystem. (B) Stable vegetation patch in drylands due to multiple stationary fronts. Parameters: $p=6.2<p_M$  (C) Coexistence states exist beyond tipping point in drylands as $p$ is changed slowly over time. Parameters: $u=0.9$, $\delta_w=2500$, $dp/dt=-4.5\times 10^{-6}$. Other parameters in A-C are the same as Fig. 4. (D) The forest state invades the savanna state in the savanna-forest system $(\mu= \mu_0)$.  (E) Stable patterns of alternate savanna(s) and forest (f) patches due to (environmental) spatial heterogeneity of water availability, expressed as variation in the forest growth rate, $\mu=\mu_0+ 0.025 \sin (0.16x)$, where $x$ is the 1D spatial coordinate. Parameters: $\mu_0=0.46>\mu_M$, other parameters are the same as Fig 4.} }
 \label{Fig5}
 \end{figure}
\begin{figure}[H]
 \begin{center}
{\includegraphics[width=1\textwidth]{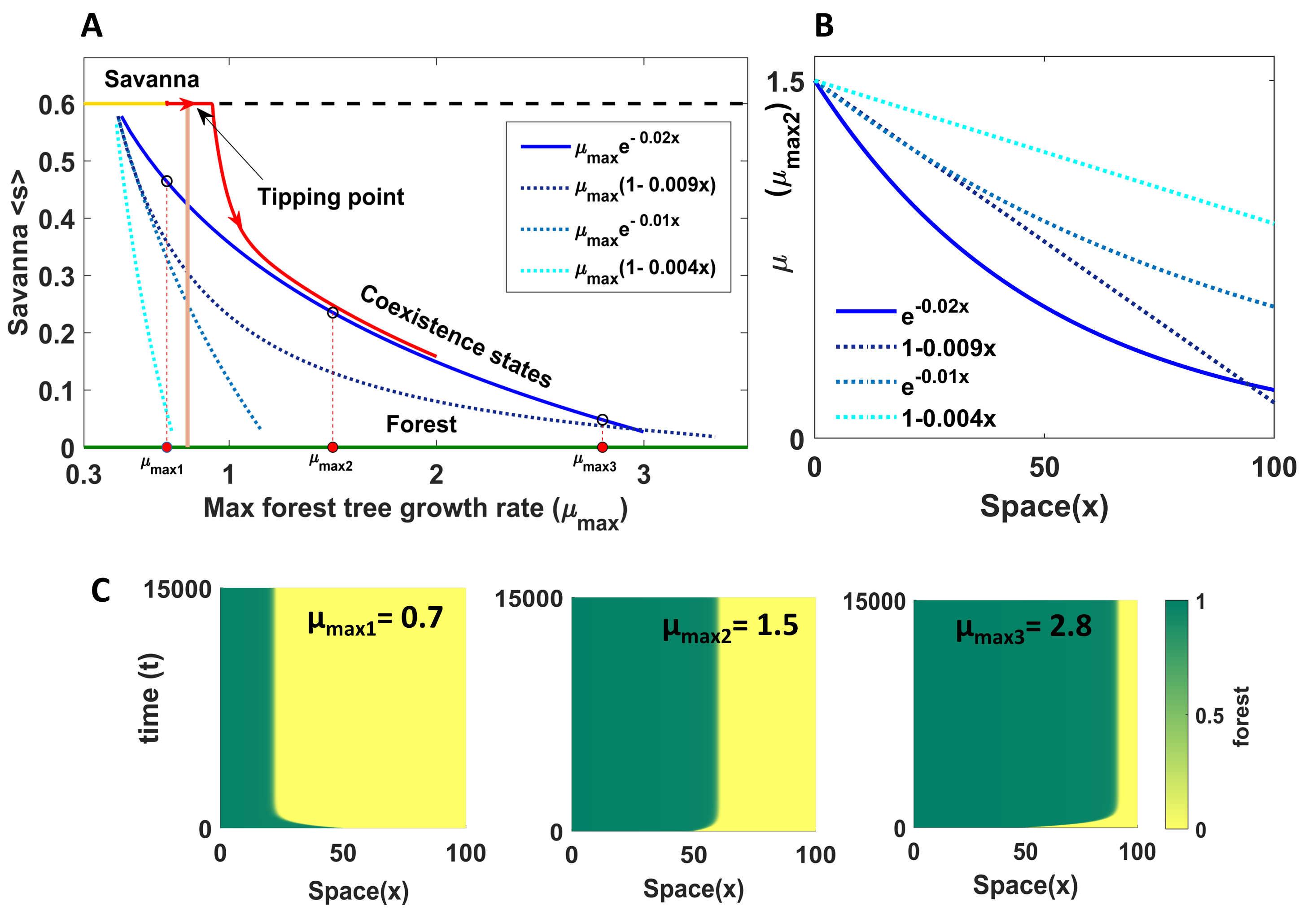}}
 \end{center}
 \caption{{\small Tipping to intermediate coexistence states when forest growth rate varies across the landscape, i.e., $\mu=\mu_{max}$ f(x). (A) Mean biomass of savanna vegetation with respect to changing $\mu_{max}$. The green  horizontal line indicates forest-only (thus no savanna biomass) while the yellow line indicates savanna only. The dashed black lines represent the unstable states. The vertical solid line denotes the tipping point of the non-spatial system. The red lines denote time series simulations with savanna state as initial condition and a gradual increase in the bifurcation parameter, i.e., $d \mu_{max}$/dt=0.0002. The blue lines indicate the intermediate coexistence states, where part of the landscape is covered with savanna and the rest with forest, for the different spatial heterogeneities, as demonstrated in (B). The circles marked on the solid blue line indicate the three values of $\mu_{max}$ for which the coexistence states are demonstrated in (C). The coexistence state for $\mu_{max1}=0.7$ indicates multistability of the ecosystem; the coexistence states for $\mu_{max2}=1.5$ and $\mu_{max3}=2.8$ are far beyond the tipping point in the non-spatial system in Fig. 2. Other parameter values are the same as in Fig. 4.  where part of the landscape is covered with savanna and the rest with forest.   } }
 \label{Fig6}
 \end{figure}

It is common in modelling studies to assume that the dynamics arising from interactions between the system components remain the same throughout the whole domain, However, in real ecosystems, local conditions can vary across the landscape with the spatial location. For example, water availability could vary quite a bit across a landscape, due to e.g. variations in topography and/or soil types. To check for stable coexistence states in the savanna-forest model, we incorporate such spatial heterogeneity of environmental conditions by assuming that the water availability determines variations in the forest growth rate, $\mu$, across the domain. We explore the consequences for two types of heterogeneity: i) spatial variation arising due to microtopography realized by considering small changes of $\mu$ around a mean, i.e., around five percent relative change in the forest tree growth rate (Fig. \ref{Fig5}E) ii) environmental gradient across a landscape realized by considering that $\mu$ decreases monotonically across the domain (Fig. \ref{Fig6}).

For the first case, we include sinusoidal variations in the forest growth rate, $\mu$, across the domain, and simulate over a one dimensional spatial domain with $\mu(x)= 0.46 + 0.025 \sin (0.16x)$, where $x$ is a 1D dimensionless spatial coordinate. If initially there is a localized domain of savanna in a homogeneous forest landscape, the whole landscape would be covered by forest in absence of the spatial heterogeneity (e.g., for $\mu$=0.46 in Fig. 5D), while in its presence, we see stable alternate patches or “coexistence states” of savanna and forest (Fig \ref{Fig5}E) . Incorporation of (relatively small) spatial heterogeneity made the propagating fronts stationary. Intuitively, this can be understood by the fact that the spatial variation leads to the environmental parameter, $\mu$, oscillating between both sides of the Maxwell point thus not allowing the fronts to completely invade the landscape resulting in stationary fronts. So, for environmental condition where forest could invade the savanna, spatial heterogeneity can lead to stable coexistence states, thus making the savanna more resilient. This could potentially explain the savanna-forest mosaics often observed at the transition zone between the two biomes \citep{charles2015functional, charles2018steal}. These findings are in accordance to earlier studies where spatial heterogeneity in environmental conditions has been shown to explain stable savanna-forest boundaries \citep{goel2020dispersal, wuyts2017amazonian, wuyts2019tropical}. So, the non-spatial model, which exhibits tipping as a result of the system being pushed to the other alternative state (N-tipping), when extended with spatial effects shows either gradual invasion of one state into another or stable coexistence of forest and savanna states with boundaries separating the two states. Note that the dryland model also exhibits similar behaviour in the presence of spatial environmental heterogeneity (see Supplementary, Section D) although as we demonstrated in the previous section, in that case, it is not a prerequisite for stable coexistence states to exist.

To explore the impact of the second type of heterogeneity, we assume that the forest tree growth rate is at its maximum at one end of the landscape $(x=0)$, and it decreases monotonically along the landscape with a minimum at the other end $(x=L)$. We choose both linear and exponential functions for $\mu(x)=\mu_{max}f(x)$ (see Fig 6). If initially a part of the landscape is covered with savanna and the rest with forest, we observe that such spatial variation of $\mu$ also gives rise to coexistence states which can remain stable for environmental conditions much beyond the tipping point predicted by the non-spatial model. This can be explained by the fact that the environmental gradient makes a part of the landscape favourable for savanna even in parameter regimes where the forest should have otherwise completely invaded. Moreover, on increasing the parameter $\mu_{max}$, instead of tipping from the savanna to the forest state, the system now switches to an intermediate stable coexistence state. On changing $\mu_{max}$ further, the system gradually changes along this intermediate stable state thus making the process less critical and more easily reversible. This is in agreement with what was shown earlier by \cite{bastiaansen2022fragmented}, though only for one component models. Further, we demonstrate that shape of the spatial gradient does not qualitatively change this system behavior. However, a larger spatial gradient of the parameter (environmental condition) results in a larger parameter regime till which the coexistence states are able to exist. So, the spatial processes in the ecosystems combined with environmental heterogeneity makes them much more resilient than that predicted by non-spatial models.

\begin{table}[H]
\tabcolsep 16 pt
\centering
\caption{\small{Overview of the different tipping evasion mechanism relevant at the two ecosystems, drylands and the savanna-forest boundary. (\ding{51}) means that our models demonstrate the mechanism while ({\large X}) means that it is absent.\\} }
\begin{tabular}{|P{4 cm}| P{2.7 cm}|P{2.7 cm}| P{4cm}|}
\hline
\textbf{\centering Mechanism} & \textbf{Drylands} &\textbf{Savanna-forest} & \textbf{Reference}  \\
& $F_v<0,$ $G_w>0$ & $H_f<0,$ $I_s<0$ & \\
\hline
& & & \\

Turing before tipping & \checkmark & {\LARGE X} & \cite{siteur2014beyond, rietkerk2021evasion}\\

& & & \\

Gradual transition via fronts & \checkmark & \checkmark & \cite{bel2012gradual, zelnik2018regime} \\

& & & \\

Stable coexistence state
without spatial heterogeneity & \checkmark & {\LARGE X} &  \cite{zelnik2018regime,jaibi2020existence}\\

& & & \\

Stable coexistence state
with spatial heterogeneity & \checkmark & \checkmark &  \cite{bastiaansen2019stable, goel2020dispersal,van2015resilience}\\

& & & \\

Front instability leading to 2D fingering patterns & \checkmark & {\LARGE X} &  \cite{fernandez2019front,carter2023criteria}\\

\hline
\end{tabular}

\label{table-1}
\end{table}

\subsection*{Front instabilities}

Although till now we only discussed front dynamics in one spatial dimension, even richer behavior can be observed when extending in two spatial dimensions. When the fronts are not stationary, local processes that occur at the front zone determine the nature of the global transitions. Boundaries or fronts separating different ecosystem states are spatial structures that can go through instabilities much like uniform states can go through Turing instabilities in response to perturbations. When that happens, the spatial interface between alternative states or coexistence states can deform and self-organize giving rise to so called “finger-like patterns”. Similar patterns of coexistence states arising out of front invasion have been studied in the past in the context of predator-prey models, including three species Lotka-Volterra models \citep{mimura2015dynamic, petrovskii2005patterns, petrovskii2002allee}, models of bacterial growth \citep{giverso2015branching} and dryland ecosystem models \citep{fernandez2019front}. In a recent study by \cite{carter2023criteria}, the authors established a precise mathematical criterion for the instability of fronts in two-component reaction diffusion models in two spatial dimensions. While the precise criterion is too technical to relate to the overall ecosystem functioning, it is possible to give the following guideline: provided a front exists, it will undergo an instability when the faster diffusing component (water, in case of the dryland model) has a positive effect on the slower diffusing component (vegetation) and the latter has a negative effect on the former. In the dryland ecosystem, which follows an activator-substrate mechanism, this condition is satisfied as $F_v<0$ and $G_w>0$. So, fronts between stable homogeneous states are typically unstable when the ratio of the diffusion coefficients of the components is suitably large, a reasonable assumption as water diffuses much faster than vegetation. This instability can lead to regular pattern formation: one observes “finger-like patterns” emerging from the interface in such models (Figure 7) which can self-organize into labyrinthine structures resembling Turing patterns. Ecologically such pattern formation is of potential profound significance: while one ecosystem state invades the other, simultaneously the latter is able to invade the former. This could even lead to reverse transitions of biomes, for example reversing desertification \citep{fernandez2019front}. However, in the savanna-forest model, which does not satisfy the above condition ($H_f<0, I_s<0$), we were unable to find any front instability, in spite of extensive numerical exploration. This indicates that interfaces in such models may in fact be stable also in two spatial dimensions so that no two-dimensional spatial patterns – e.g. fingers or labyrinths – will be formed. 





\begin{figure}[H]
 \begin{center}
{\includegraphics[width=1\textwidth]{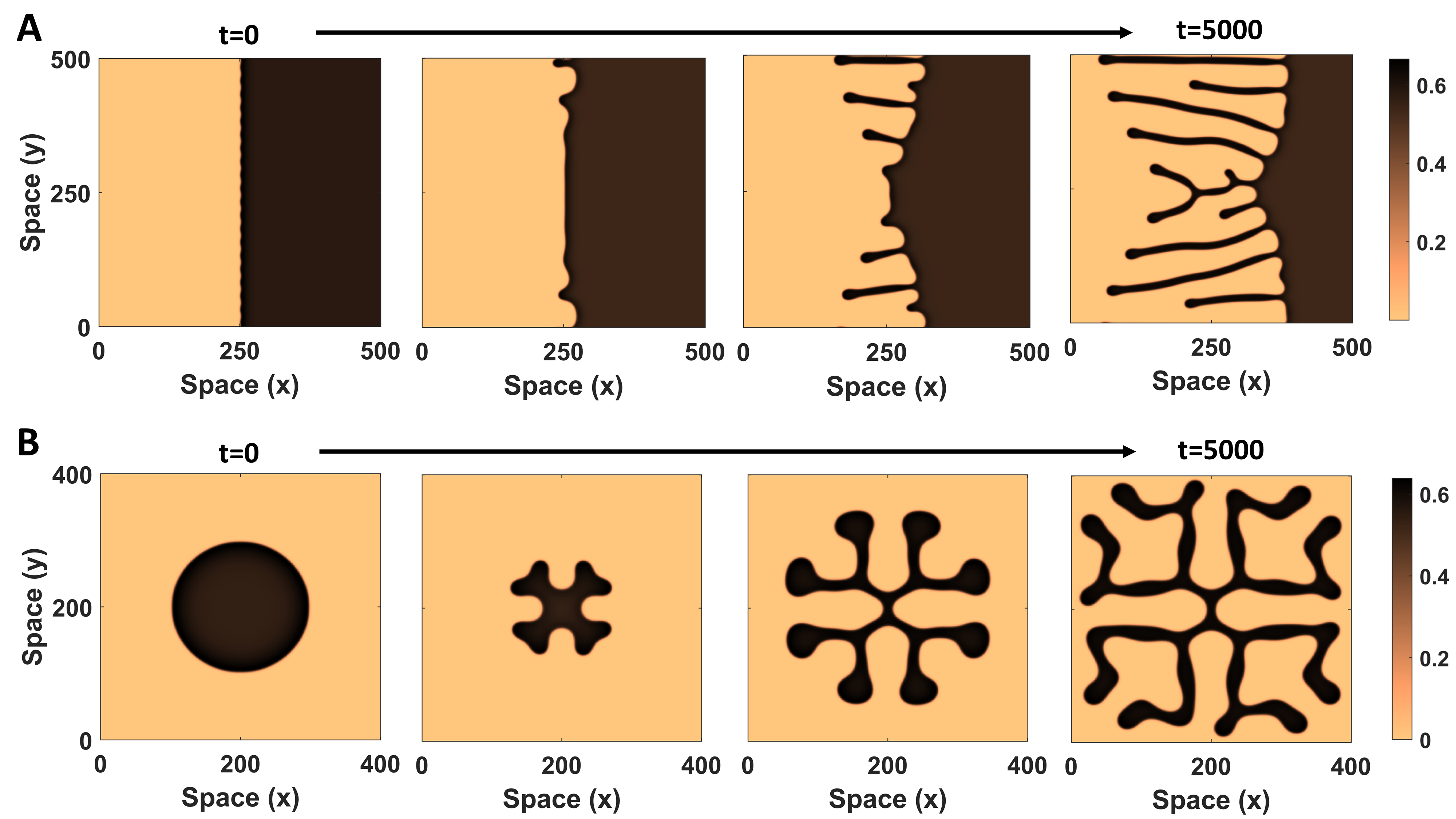}}
 \end{center}
 \caption{{\small Front instabilities in the dryland model. Vegetation density is indicated by the color bar. Dark brown represents the vegetated state and light brown represents barren state. Simulations are initialized with vegetation covering (A) half of the landscape ($\delta_w=100$)(B) a circular patch within the landscape ($\delta_w=400$), and snapshots at different times are shown from left to right. Parameter values: p=6.2; the other parameters are the same as in Fig 2.  } }
 \label{Fig7}
 \end{figure}

\section*{Discussion and future work} 

Our study highlights the role of spatial processes in ecosystem response to global change and showcases the distinct mechanisms underlying pattern formation that are relevant when there exist alternative stable states. Both humid savanna-forest transition zone and dryland ecosystems, which are used for the purpose of this study, are believed to be prone to tipping. So, we use models of these ecosystems to synthesize the important mechanisms that could lead to intermediate states and tipping evasion when relatively simple spatial dynamics are incorporated. Since the nature of the ecological interactions driving the dynamics are different in the two systems, our synthesis indicates a link between the tipping evasion mechanisms and the key ecological interactions.

In both the drylands and humid savanna-forest boundary, much before the ecosystem reaches its tipping point, if there is a localized disturbance, there is a gradual invasion of one state into the other instead of (N-)tipping. Further, in drylands, multiple localized disturbances in such environmental conditions may also form stable patterns, i.e., vegetation and barren land may form stable coexistence states, which can persist much beyond the tipping point. However, in the savanna-forest model, such stable coexistence can only be observed when environmental conditions such as topography or soil type or annual rainfall are spatially heterogeneous. Interestingly, in this case, the type of heterogeneity determines whether or not these coexistence states can persist beyond the tipping point. When they indeed persist (under large enough heterogeneity), instead of a complete transition from savanna to tropical forest, only a part of the landscape tips followed by gradual transitions between the stable savanna-forest coexistence states.

Drylands, which functions as an activator-substrate systems, can also demonstrate spatial patterns due to the onset of Turing instability or front instability before the tipping point is reached, which was not observed in the savanna-forest model. These, in fact, are also additional mechanisms by which the ecosystem can evade tipping without the need for underlying spatial heterogeneity. After the onset of Turing patterns, the ecosystem is able to adapt to patterns of increasingly larger wavelengths on increasing stress, thus undergoing small transitions instead of one large critical transition or tipping. At a much larger annual rainfall, far from the point where Turing instability sets in, front instability may occur which may result in finger-like patterns of alternative stable states situated next to each other. These patterns eventually in the long run self-organize into labyrinth like structures resembling Turing patterns. 

So, the pathway of tipping evasion relevant for drylands and humid savanna-forest transition zone might be different. Table 1 provides an overview of the different mechanisms that we discussed in this synthesis and whether they are relevant to the two systems that we studied. While intrinsic mechanisms make drylands resilient, humid savannas may have to rely on spatial heterogeneity of the environment to avoid tipping. The differences may be attributed to the key mechanisms at play in the two biomes which are also reflected in the model structure. While both ecosystems exhibit positive feedback, the mechanism which drives this positive feedback is quite distinct. In the drylands, vegetation has a self-enhancing effect through increasing water infiltration. In fact, it follows an activator-substrate mechanism $(F_v<0, G_w>0)$ while at the savanna-forest boundary, the negative effect of the two vegetation types on each other is the dominant factor $(H_f<0, I_f<0)$. This provides a fresh impetus to further investigate the connection between ecological interactions and tipping evasion in other ecosystems.

The synthesis demonstrates that it is important to appreciate the spatial complexity of ecosystems and the key interactions driving the ecosystem dynamics while predicting its response and resilience to global change. Although the models used in this paper provide meaningful insights, it is important to acknowledge that they are quite simple, and realistic spatial effects are more complicated, possibly leading to even more intricate dynamics in the real world. Future research should not only strive towards developing the theory of tipping evasion, but also towards identifying spatial patterns in real ecosystems which are believed to be tipping-prone and analyzing data from observations and remote sensing. This can help understand the mechanisms behind such pattern formation and their resilience to changing climatic conditions which will allow better understanding of ecosystem response to global change.

\section*{Acknowledgments}

S.B. would like to acknowledge European Union’s Horizon 2020 research and innovation programme under the Marie Sklodowska–Curie grant agreement no 101025056 (project ‘SpatialSAVE’) for funding this research. S.B. also acknowledges his present funidng via the Foundations and Applications of Emergence (FAEME) programme at the Dutch Institute for Emergent Phenomena (DIEP), University of Amsterdam. P.C. was supported by the National Science Foundation through the award DMS-2105816. M.B. was supported by the Italian National Biodiversity Future Center (NBFC): National Recovery and Resilience Plan (NRRP), Mission 4 Component 2 Investment 1.4 of the Italian Ministry of University and Research; funded by the European Union – NextGenerationEU (Project code CN-00000033) and a Visiting Professor grant of the Centre for Complex System Studies (CCSS) of the Utrecht University. The research of M.R. and A.D. is supported by the European Research Council (ERC-Synergy project RESILIENCE, proposal nr. 101071417) and by the Dutch Research Council (NWO ‘Resilience in complex systems through adaptive spatial pattern formation’, project nr. OCENW.M20.169). S.B. and M.B. would also like thank Antoine Guines for stimulating them with his master project which led to the exploration shown in the Supplementary Sec E.


\section*{Statement of Authorship}

SB, MB, AD, MR contributed to the conceptualization and methodology. SB, PC and RB contributed to the formal analysis. SB prepared the visual materials. SB wrote the original draft. All authors contributed to reviewing and editing.



\bibliographystyle{amnatnat}
\bibliography{article}

\newpage

\end{document}